\documentclass[notitlepage,nofootinbib]{revtex4-1}
\usepackage[latin9]{inputenc}
\usepackage{amsmath}
\usepackage{graphicx}
\usepackage{comment}
\usepackage{xcolor}
\usepackage{hyperref}
\usepackage[normalem]{ulem}
\usepackage{comment}

\begin{document}
\title{The distribution of Bayes' factor}
\author{Luca~Amendola$^{1,a}$, Vrund Patel$^{1,a}$, Ziad Sakr$^{1,}$\footnote{L.A., V.P., Z.S. contributed equally to this work.}}
\affiliation{$^{1}$Institut für Theoretische Physik, Universit\"at Heidelberg,
Philosophenweg 16, 69120 Heidelberg, Germany
}
\author{}
\author{Elena Sellentin$^{2,3}$, Kevin Wolz$^4$}
\affiliation{$^2$ Mathematical Institute, Leiden University, Gorleaus Building, Einsteinweg 55, NL-2333 CA Leiden, The Netherlands}
\affiliation{$^3$ Leiden Observatory, Leiden University, Gorleaus Building, Einsteinweg 55, NL-2333 CA Leiden, The Netherlands}
\affiliation{$^{4}$Department of Physics, University of Oxford, Denys Wilkinson Building, Keble Road, Oxford OX1 3RH, United Kingdom}
\begin{abstract}

The ratio of Bayesian evidences is a popular tool in cosmology to compare different models. There are however several issues with this method: Bayes' factor depends on the prior even in the limit of non-informative priors, and the Jeffreys scale, used to assess the test, is arbitrary. 
Moreover, the standard use of Bayes' factor is  often criticized for being unable to reject models. In this paper, we address these shortcoming by promoting  evidence ratios to frequentist statistics and deriving their sampling distributions. By comparing the evidence ratios to their sampling distributions, poor fitting models can now be rejected.   
Our method additionally 
does not depend on the prior in the limit of very weak priors, thereby  safeguarding the experimenter against premature rejection of a theory with a  uninformative prior, and replaces the arbitrary Jeffreys scale by probability thresholds for rejection. We provide analytical solutions for some simplified cases (Gaussian data, linear parameters, and nested models), and we apply the method to cosmological supernovae Ia data.
We dub our method the FB method, for Frequentist-Bayesian.
\end{abstract}
\maketitle

\section{Introduction}

Modern cosmology is  entering an era where large amounts of data will be forthcoming from cosmic microwave background, large scale structure, distance indicators, and gravitational waves. This requires a thorough understanding of the employed statistical methodology in order to make robust inferences from the data. Often investigations are performed in terms of a Bayesian study where the  analysis of competing models is typically performed in three steps: first, the  confidence region of a model is produced; secondly, two or more competing models are compared via Bayes' factors; and third, the quality of fit of the models is assessed via a simple $\chi^2$ test \cite{2023arXiv230315401D,Reeves:2022aoi,Alonso:2023guh,2023arXiv231019899M,2023arXiv231109936K,Yang:2022klj,Holm:2023laa,2023arXiv230917151S,Colgain:2023bge,Sakr:2021jya,Serra_2007,2008ApJ...675....1K,2002PhRvD..65d3506J,Shafer:2015kda,Heavens:2017hkr} 
These three tools are based on heterogeneous assumptions: for instance, the $\chi^2$ quality of fit is a frequentist test that does not take into account priors, and the parameter confidence region cannot tell, {\it per se}, if an alternative point hypothesis outside this region is excluded, until we perform a specific model comparison. Moreover, concerning   Bayes' factor test limitations, its interpretation relies on  arbitrary criteria to map its value onto the commonly used the Jeffreys \cite{Jeffreys:1961,2008arXiv0804.3173R}, or Raftery-Kass scale \cite{Kass:1995loi}. Another important issue is that the result depends on the considered priors, even in the weak limit or in the case where an informative flat prior is used, since the evidence is averaged over the prior volume which can impact differently each model. Finally, the Bayesian comparison cannot reject models: it can only determine which model is relatively better, but not whether either model is a good fit to the data on its own. 

Several other issues have been discussed in the literature. For instance, Ref. \cite{Efstathiou:2008ed} showed that computing Bayesian evidence using cosmological data could yield, wrongly, a statistically insignificant preference for a model over a physically poorly motivated alternative model. Ref. \cite{Nesseris:2012cq} showed that when comparing nested models, the Jeffreys scale can fail to penalize extra degrees of freedom when mock data are generated using the simpler model. 
Ref. \cite{Jenkins:2011va}  noted that the performance of Bayes' factor test is strongly affected by the signal to noise ratio in
the data along with the non comprehensiveness of the models under consideration.

There have been some attempts, still within the Bayesian framework,
to remedy for some of these limitations, in particular for the models comparison or rejection issue, by proposing  alternatives  such as the Akaike information criterion (AIC) test, the Bayesian information criterion (BIC) test or the deviance information criterion (DIC). 
In that regard, \cite{Rezaei:2021qpq} performed a comparison of the above tests with  Bayes' factor using cosmological data and found 
different levels of agreement between the outcome  for certain models and probes.
Ref.~\cite{10.1214/ss/1009212519}, or \cite{Parkinson:2013hzz} for a review, proposed the Bayesian Model Averaging method where each individual posterior, generated under the assumption of a particular model, is weighted by the model likelihood and then combined with the other posteriors to give a model-independent posterior. This  method is especially useful in case the data are not strong enough to distinguish decisively between the models. Ref. \cite{Kunz:2006mc} introduced a statistical measure of the effective model complexity that assesses how many effective
parameters a set of data can support and demonstrated that it can be used as a useful complement to  Bayes' factor in model selection questions. Ref. \cite{Raveri:2018wln} develops a full set of estimators of internal and mutual agreement and disagreement
between datasets whose strengths complement each other. This allows to take into account the effect of prior information and the statistical significance of discrepancies and unknown systematics.  Ref. \cite{2023JCAP...11..048H} studied the sampling properties of Bayesian evidences as a frequentist statistic when compressing the data from its original size down to the number of free parameters. Finally, \cite{Koo:2021suo} proposed a method in which the likelihood distribution of the difference in $\chi^2$ between a smoothed function and that of the best-fit of the model being tested is calculated. They then  use the values that enclose 95\% and 99\% of  the likelihood volume as criterion to test whether data support a given model, independent of how well other models perform, though they do not assess the effect of the prior on the final decision.

In this work we  propose instead the use of  Bayes' factor as a statistic to perform a frequentist $p$-value test. Moreover, we implement a frequentist version of Bayesian model comparison by using the odds ratio, i.e. the ratio of probabilities, between two competing models. These two complementary tools, $p$-value and frequentist model comparison, have interesting properties: they  are entirely based on the same statistic; they generally depend on the prior but not in the limit of weak priors; they can assess both the quality of the fit and the relative probability. 
 Assuming two competing models $A$ and $B$, the final result can give one of the following answers: the data are compatible with $A$ but not with $B$; they are compatible with $B$ but not with $A$; they are compatible with both; they are incompatible with both. Finally, regardless of the above, the model comparison test can tell us whether $A$ is favored over $B$ or viceversa. We illustrate the whole procedure with two complementary types of plots that  provide a full analysis of the problem: one comparing the Bayes' factor distributions with the real data Bayes' factor, and another one comparing the model comparison odds ratio to the real data odds ratio. 
 
Perhaps the first suggestion to perform a frequentist test with Bayes' factor is in Ref. \cite{Good1957} (see review in \cite{Good1992}). A study of the frequentist distribution of  Internal Robustness (\cite{Amendola:2012wc}), which is formed with ratio of evidences, was carried out in \cite{Kevinthesis}.
In Ref.~\cite{Passenger:2024piv}, a Bayesian evidence was also considered as a statistic in the form of a $p$-value,  to assess whether observed exceptional gravitational-wave events are consistent with extremal simulated events.  Ref.~\cite{Joachimi:2021ffv} studied the impact on model comparison if evidences are to be interpreted based on the ensemble of possible observations, and derived the mean and variance of Bayes's factor.
Several further works studied various aspects of this concept, see for instance \cite{Castro:2014oja, 2021arXiv211015625F}.

In this paper, the approach is quite close to Ref.~\cite{Keeley:2021dmx}. They generated a distribution of evidences from datasets assuming a model and showed that the model is ruled out if its evidence  lies outside that distribution. 
Here, in addition, we will assess the impact of the prior on the final decision and put forward a frequentist analog of the Bayesian model comparison. We also provide the analytical Bayes' factor distribution for  correlated Gaussian data with linear nested models (thereby extending  \cite{2023JCAP...11..048H} who studied model rejection under data compression). In this paper, we apply our method, to be denoted   as FB method, to a simple polynomial fit and to supernovae Ia data.

\section{Bayes' factor for Gaussian data and linear models}
\label{sec:BayesGaussLin}

Bayes' factor  is used in general to compare two competing models, where the fiducial model is compared to a new one by computing the ratio $b$  defined as 
\begin{equation}b= \frac{E({ D}|{ M}_A)}{E({ D}|{ M}_B)}, 
\end{equation}
where $E$ is the evidence 
\begin{equation} E({ D}|{ M}) = \int d \theta\ {\cal L}({ D}| \theta,{ M})\, {\cal P}( \theta,{ M})\,,
\end{equation}
with ${\cal L}(D|M_{A , B})$
the probability of the data $D$ given model $M_{A , B}$ with parameters $\theta_\alpha$ and a prior ${\cal P}$. Model $M_A$ is favored over model $B$ if $b >  1$, and $M_B$ is favored otherwise. 

In this and the next section we will find the analytical distribution of Bayes' factor in a particular, but
important, case, namely Gaussian data with linear nested models. In this section we start  with reviewing the formalism and the general problem 
{\bf (see \cite{Trotta:2008qt} for a review})
.

Linear models can be described by a sum  $f_{i}=\theta_{\alpha}g_{\alpha i}$
where $\theta$ is the parameter vector  and $g_{\alpha i}$ denotes the linear mapping between the $\alpha$-th model parameter and the $i$th datapoint. For example, the parameters may act as the linear coefficients of a polynomial in a deterministic variable $x$, $f_i=\theta_{0}+\theta_{1}x_i+\theta_{2}x_i^{2}+...$. We use Greek subscripts
referring to parameters, lowercase Latin subscripts referring to data,  and capital
Latin subscripts to refer to models. Moreover, we adopt the convention that repeated indexes are summed over.
The likelihood for $N$ data points $d_i$ is 
\begin{equation}\label{eq:genlik}
   {\mathcal L}=(2\pi)^{-N/2}|\Sigma|^{-1/2} \exp -\frac{1}{2}(d_i-f_i)\Sigma^{-1}_{ij}(d_j-f_j)
\end{equation}
where $\Sigma$ is the data covariance matrix.
We also assume a Gaussian prior with mean vector $\tilde\theta$
\begin{equation}
   {\mathcal P}=(2\pi)^{-N_M/2}|P|^{1/2} \exp -\frac{1}{2}(\theta_\alpha-\tilde\theta_\alpha)P_{\alpha\beta}(\theta_\beta-\tilde\theta_\beta)
\end{equation}
where $N_M$ is the number of parameters for model M. The posterior is therefore $\mathcal{LP}/E$, where $E$ is the evidence. In the limit of weak  prior, $P_{\alpha\beta}\to 0$. We will often adopt this limit to obtain simpler expressions. Frequently used symbols are listed in Table \ref{tab:symbols}.

 The bestfit (maximum likelihood)
parameters are found by imposing $\partial {\mathcal L}/\partial\theta_\alpha=0 $, i.e. solving the equation
\begin{equation}\label{eq:bfeq}
    g_{\alpha i}\Sigma^{-1}_{ij}(d_i-\hat\theta_\beta g_{\beta j})=0
\end{equation}
from which
$\hat{\theta}_\beta=D_\alpha L^{-1}_{\alpha\beta}$ where $L_{\alpha\beta}=g_{\alpha i}\Sigma_{ij}^{-1}g_{\beta j}$
is the Fisher matrix,  and
$D_{\alpha}=d_{i}\Sigma_{ij}^{-1}g_{\alpha j}$. 
In the limit of weak prior, $\hat\theta $ coincides with the maximum of the posterior, and therefore gives the Bayesian best fit solution. For general Gaussian priors, the posterior maximum is 
\begin{equation}\label{equ:theta}
    \theta^*=(\hat\theta L+\tilde\theta P)F^{-1}\,.
\end{equation}

For linear models and Gaussian data, the maximum likelihood parameters are Gaussian random variables and the Fisher matrix represents their exact inverse covariance. In Sec. \ref{sec:Fisher} we briefly discuss the Fisher approximation  for non-linear models.

We  define the matrix $G$  as $g_{\alpha i}L_{\alpha\beta}^{-1}g_{\beta j}\equiv G_{ij}$. In index-free notation, $\hat\theta^T=D^TL^{-1}$, $D^T=d^T\Sigma^{-1}g^T$, $G=g^T L^{-1}g$ and $L=g\Sigma^{-1}g^T$. We will use both indexes and index-free notation, whichever is clearer in each case.
It follows 
\begin{equation}\label{equ:f}
f=\hat{\theta}^Tg= D^TL^{-1}g=d^T\Sigma^{-1}g^TL^{-1}g=d^T\Sigma^{-1}G
\end{equation}
We see that $C\equiv\Sigma^{-1}G$ is the matrix that projects the
information from $d$ to the theory best-fit $g^T\hat{\theta}$. (Note
that $C^{2}=C,$ so it's indeed a projector; it's also a symmetric
matrix). Note also that 
\begin{align}\label{eq:cproj}
\mathrm{Tr}\, C={\mathrm Tr}(\Sigma^{-1}
G)={\mathrm Tr}(\Sigma^{-1}
g^TL^{-1}g)={\mathrm Tr}(g\Sigma^{-1}
g^TL^{-1})={\mathrm Tr}(LL^{-1})=N_M
\end{align}
where $N_{M}$ is the number of parameters of model $M$. Then on the
best fit one has 
\begin{align}
\hat{\chi}^{2} & =(d^T-\hat{\theta}^Tg)\Sigma^{-1}(d-g^T\hat{\theta})\label{eq:chihat-2}\\
 & =d^T\Sigma^{-1}(I-C)d
\end{align}
 We now rewrite the likelihood as a function of the parameters, by writing $f_i=\hat\theta_\alpha g_{\alpha i}+(\theta_\alpha-\hat\theta_\alpha)g_{\alpha i}$ and using the fact that 
\begin{align}
\frac{1}{2}(d_i-f_i)\Sigma^{-1}_{ij}(d_j-f_j)&=
\frac{1}{2}(d_i-\hat\theta_\alpha g_{\alpha i})\Sigma^{-1}_{ij}(d_j-\hat\theta_\beta g_{\beta i})-(\theta_\alpha-\hat\theta_\alpha)g_{\alpha i}\Sigma^{-1}_{ij}(d_j-\hat\theta_\beta g_{\beta i})+\frac{1}{2}(\theta_\alpha -\hat\theta_\alpha) g_{\alpha i}\Sigma^{-1}_{ij}g_{\beta j}(\theta_\beta -\hat\theta_\beta) \nonumber\\
&=\frac{1}{2}\hat\chi^2+\frac{1}{2}(\theta_\alpha -\hat\theta_\alpha) L_{\alpha\beta}(\theta_\beta -\hat\theta_\beta)&
\end{align}
since $(\theta_\alpha-\hat\theta_\alpha) g_{\alpha i}\Sigma^{-1}_{ij}(d_j-\hat\theta_\beta g_{\beta i})=0$ on account of Eq. (\ref{eq:bfeq}). Therefore
\begin{equation}
   {\mathcal L}=(2\pi)^{-N/2}|\Sigma|^{-1/2} e^{-\frac{1}{2}\hat\chi^2}\exp -\frac{1}{2}(\theta_\alpha-\hat\theta_\alpha)L_{\alpha\beta}(\theta_\beta-\hat\theta_\beta)
\end{equation}
Using this expression, one finds that the evidence for Gaussian data, linear model and Gaussian priors is
\begin{equation}
E=\int{\mathcal L \mathcal P}d^N\theta=(2\pi)^{-N/2}|\Sigma|^{-\frac{1}{2}}e^{-\frac{1}{2}\hat{\chi}^{2}}\frac{|P|^{1/2}}{|F|^{1/2}}\exp\left[-\frac{1}{2}(\hat{\theta}-\tilde{\theta})^TLF^{-1}P(\hat{\theta}-\tilde{\theta})\right]\,.\label{eq:evidence-1}
\end{equation}

\begin{table}

\begin{centering}
\begin{tabular}{c|c}
\hline 
symbol & meaning \tabularnewline
\hline 
\hline 
$\Sigma_{ij}$ & data covariance matrix\tabularnewline
\hline 
$d_{i}$ & data points\tabularnewline
\hline 
$g_{\alpha i}$ & fit functions  matrix (design matrix)\tabularnewline
\hline 
$P_{\alpha\beta}$ & prior precision matrix (inverse of the parameter covariance matrix)\tabularnewline
\hline 
$L_{\alpha\beta}$ & likelihood parameter Fisher matrix\tabularnewline
\hline 
$F_{\alpha\beta}=L_{\alpha\beta}+P_{\alpha\beta}$ & posterior parameter Fisher matrix\tabularnewline
\hline 
$N$ & number of data points\tabularnewline
\hline 
$N_M$ & number of parameters for model M\tabularnewline
\hline 
$\hat{\theta}_\alpha$ & vector of best-fit parameters (dependent on the data)\tabularnewline
\hline 
$\tilde{\theta}_\alpha$ & vector of prior means\tabularnewline
\hline 
$\hat{\chi}^{2}$ &  minimum $\chi^{2}$ (dependent on the data)\tabularnewline
\hline 
$\beta=2\log b$ & log of ratio of evidences $b$ (Bayes' ratio)\tabularnewline
\hline 
\end{tabular}\caption{\label{tab:symbols} Frequently used symbols.}
\par\end{centering}
\end{table}
 In the linear fit case, the data (our random variables) are only in $\hat\theta$ and in $\hat\chi^2$; all the other quantities are not random variables.

 Bayes' ratio for models $A$ (with $N_{A}$ free parameters) and $B$
(with $N_{B}>N_A$ free parameters)  defined as the ratio of evidences, becomes then \textbf{(see \cite{Lazarides:2004we,Heavens:2007ka} for similar calculation of the Bayes factor with the Fisher matrix approximation)} 
\begin{align}
b & \equiv \frac{E_{A}}{E_{B}}=\frac{|\Sigma|^{-\frac{1}{2}}e^{-\frac{1}{2}{\hat\chi}_{A}^{2}}\frac{|P_{A}|^{1/2}}{|F_{A}|^{1/2}}\exp\left[-\frac{1}{2}(\hat{\theta}_{A}-\tilde{\theta}_{A})^TL_{A}F_{A}^{-1}P_{A}(\hat{\theta}_{A}-\tilde{\theta}_{A})\right]}{|\Sigma|^{-\frac{1}{2}}e^{-\frac{1}{2}\hat{\chi_{B}}^{2}}\frac{|P_{B}|^{1/2}}{|F_{B}|^{1/2}}\exp\left[-\frac{1}{2}(\hat{\theta}_{B}-\tilde{\theta}_{B})^TL_{B}F_{B}^{-1}P_{B}(\hat{\theta}_{B}-\tilde{\theta}_{B})\right]}\\
 & =\exp\left[-\frac{1}{2}(\hat{\chi}^{2}_{A}-\hat{\chi}_{B}^{2})-\frac{1}{2}(\chi_{P,A}^{2}-\chi_{P,B}^{2})\right]\frac{|P_{A}F_{B}|^{1/2}}{|P_{B}F_{A}|^{1/2}}
\end{align}
where
\begin{equation}
\chi_{P,M}^{2}=(\hat{\theta}_{M}-\tilde{\theta}_{M})^TL_{M}F_{M}^{-1}P_{M}(\hat{\theta}_{M}-\tilde{\theta}_{M})
\end{equation}
We have  then
\begin{align}\label{eq:fullbeta}
\beta & \equiv2\ln b=\chi_{B}^{2}-\chi_{A}^{2}+\chi_{P,B}^{2}-\chi_{P,A}^{2}+Y_{B}-Y_{A}
\end{align}
  where
\begin{equation}
Y_{M}=\ln\frac{|F_{M}|}{|P_{M}|}
\end{equation}
with $M=A,B$. In Eq. (\ref{eq:fullbeta}), the first pair of terms gives the distance between the likelihood maxima, the second pair measures how close are the best-fits to the prior mean, and the third pair measures
how new data improve upon the prior. In the standard Bayesian approach, model A is preferred if $\beta$ is positive and large, that is, when it is a good fit (small $\chi^2_A$), which does not depart too much from the prior (small $\chi^2_{P,A}$), and when the gain from the prior to the final uncertainty is minimal  (small $Y_A$).
From Eq.~(\ref{eq:fullbeta}) we already intuitively see that a `robust' rejection of one model (i.e. a rejection that holds up under repeated draws of the data) depends on the relative magnitude of the $\chi^2_M$ terms with respect to the $Y_M$ terms. If $Y_M \gg \chi^2_M$, then $\beta$ will hardly scatter under repetitions of the experiment. However, $Y_M$ sets the experiments measurement precision with respect to the experiments prior range. We therefore see that an experiment that fails to strongly update the prior indeed also produces a $\beta$ that strongly scatters, and hence warrants a comparison to the sampling distribution of $\beta$ for reliable model selection.

We removed the hat from $\chi^{2}$ since from now on we interpret $\beta$ as a
frequentist statistic, i.e. as a random variable which can assume any value, while $\hat{\chi}^{2}$
was the specific value obtained with the real dataset.

For nested models, there will be a number of common parameters, and some additional  parameters for model B (we call them "common parameters" and "extra parameters", respectively, in the following).  We assume  that the prior means for the common parameters are the same for both models, and zero for the extra ones (this is not restrictive: one can always redefine the extra parameters so that when they vanish model B reduces to model A). The elements $g_{\alpha i}$ will then be the same for the common parameters, and the prior mean vector  $\tilde\theta$ for model B will be the same as for model A, plus as many zeros as the extra parameters.

 In the limit of weak prior, since
\begin{align}L_{A}F_{A}^{-1}P_{A} & =L_{A}(L_{A}+P_{A})^{-1}P_{A}\\
 & =(P_{A}^{-1}+L_{A}^{-1})^{-1}\approx P_{A}
\end{align}
vanishes, also $\chi^2_{P,M}$ vanishes, and we have
\begin{equation}\label{eq:lincombchi}
\beta={\chi}^{2}_{B}-{\chi}_{A}^{2}+Y_B -Y_A\,.
\end{equation}
In this limit, the statistic $\beta$ has the same distribution as the log of the likelihood ratio (up to the constant $Y_B -Y_A$).

 If, in the same weak prior limit, for simplicity one assumes that both $F$ and $P$ are diagonal matrices with entries $\sigma^{-2}$ and $\sigma_P^{-2}$, respectively, for each parameter, and  $\sigma_P=\alpha\sigma$, then one has $|F_M|=\sigma^{-2N}$ and $|P|=\sigma_P^{-2N}$, 
  and therefore
\begin{equation}\label{eq:limbeta}
\beta={\chi}^{2}_{B}-{\chi}_{A}^{2}+2r\log(\alpha)
\end{equation}
where $r=N_B-N_A$,
which always favors $A$ for $\alpha\to\infty$. This argument can be directly extended to any likelihood, not just Gaussian. In fact, if the prior is very broad with respect to the likelihood, then its value can be taken approximately constant over a large volume $V_M$ in parameter space, so that ${\mathcal P}\approx 1/V_M$. Then one has
\begin{equation}
    b\approx \frac{\int {\mathcal L}{\mathcal P}d^{N_A}\theta_{A} }{\int {\mathcal L}{\mathcal P}d^{N_B}\theta_{B}}\to \frac{V_B \int {\mathcal L}d^{N_A}\theta_{A} }{V_A\int {\mathcal L}d^{N_B}\theta_{B}}
\end{equation}
If models are nested, then $V_B=V_A V_e$, where $V_e$ is the volume of the extra parameters. Then again one sees that $b$ increases arbitrarily for large prior volume $V_e$.
The unappealing consequences of this behavior in a simple but extreme example are illustrated in App. \ref{sec:simple}. 

Note that the artificial suppression of the new model  arises because the prior was subjectively chosen as one that is effectively a flat prior with wide boundaries.
If the prior is  derived by sound physical or mathematical principles (see e.g. \cite{priors} who computed an uninformative prior that does not disturb the low signal-to-noise measurements of elusive neutrino physics), then of course
the prior is part of the model and no issue arises.

\section{Analytical distribution of Bayes' ratio}

Our next goal now is to answer this question: given that the data are Gaussian random variables distributed around a given model, what is the distribution of $\beta$ ? We always assume the models to be nested, i.e.  model B has a subset of  parameters that coincide with those of model A, and with  the same prior, plus a number of extra parameters 
with their corresponding prior. In other words, model A is a constrained version of model B. Moreover, without lack of generality, model B reduces to model A if the extra parameters are put to zero. This applies, for instance, to $\Lambda$CDM and $w$CDM, which is  the real data example we investigate later on, provided one takes as extra parameter $w'=w-1$.

 It is important to realize that the random variables (the data) enter quadratically both in $\chi^2_{A/B}$ and in $ \chi^2_{P,B/A}$. A  positive-defined quadratic form of 
  Gaussian variables $x_i M_{ij}x_j$, as $\beta$ in Eq. (\ref{eq:lincombchi}), is distributed as a $\chi^2$ variable.  
There are however three types of $\chi^2$ variables: standard, non-central, and generalized. If the Gaussian variables $x$ have zero mean, the $\chi^2$ distribution is the standard one, defined uniquely by the degrees of freedom (dof) $\nu$. If the Gaussian variables have mean $\mu_i$  such that 
\begin{equation}
    \lambda=\mu_i M_{ij}\mu_j
\end{equation}
then they are non-central $\chi^2$ variables, defined by $\nu$ and $\lambda$. Finally,  a linear combination of non-central $\chi^2$ variables is distributed as a generalized $\chi^2$ variable. This distribution is defined by the $\nu,\lambda$ of each $\chi^2$ variables, and by their coefficients of the linear combination. More details on these distributions will be presented below.

 As we will see, the standard case applies to nested models with weak priors when assuming A to be true and the prior mean to coincide with the best fit values. The non-central case applies to nested models with weak priors when the prior mean does not coincide with the best fit values, or with any prior but the two models differ by just one parameter. The generalized case, finally, applies in all other situations. 
We need also consider variables that are shifted and scaled versions of $\chi^2$ variables, i.e. $z\equiv \alpha\chi^2+a$, and their distributions are trivially obtained by shifting and scaling the argument of the original $\chi^2$ distributions.

 We need then to show that
the combination $(\chi_{B}^{2}+\chi_{P,B}^{2})-(\chi_{A}^{2}+\chi_{P,A}^{2})$,
which are the terms that depend on data, is a quadratic form and find out its distribution.
For each term we have 
(we
temporarily suppress here the subscript A or B) 
\begin{equation}\label{eq:chichip}
\chi^{2}+\chi_{P}^{2}  =\Xi +\Pi
\end{equation}
where 
\begin{align}
\Xi &\equiv(d_{m}-u_{m})T_{mk}(d_{k}-u_{k})\\
\Pi & \equiv -\tilde{\theta}_{\gamma}(W_{m\gamma}T_{mk}W_{k\alpha}-L_{\gamma\beta}F_{\beta\delta}^{-1}P_{\delta\alpha})\tilde{\theta}_{\alpha}\\
u_{k} & =T_{kn}^{-1}\Sigma_{n\ell}^{-1}g_{\alpha\ell}(F^{-1}P)_{\alpha\gamma}\tilde{\theta}_{\gamma}=W_{k\gamma}\tilde{\theta}_{\gamma}\\
T & =\Sigma^{-1}(I-g^{T}F^{-1}g\Sigma^{-1})
\end{align}
However, one can show that $\Pi=0$.
It turns out that 
\begin{equation}\label{eq:t-1}
T^{-1}=\Sigma+g^{T}P^{-1}g
\end{equation}
and therefore 
\begin{equation}\label{eq:u}
u =g^{T}\tilde{\theta}
\end{equation}
i.e., $W=g$. 
Since $\Xi$ is a quadratic form of Gaussian variables, it is distributed as one of the three types of $\chi^2$ variable.

 Let us denote with $\mu_{m}$ the expected value of $d_{m}-u_{m}$
\begin{equation}
\mu=\langle d -u \rangle=g^{T}(\hat{\theta}-\tilde{\theta})
\end{equation}
We also have $\langle d \rangle=f=\hat\theta g$ and, by definition of covariance matrix,
\begin{equation}
    \langle (d-g^T\hat\theta )(d^T-\hat\theta^T g)\rangle=\Sigma
\end{equation}
so that
\begin{align}
    \langle\Xi\rangle&=\langle(d-g^{T}\hat{\theta})+\mu)^T T((d-g^{T}\hat{\theta})+\mu)\rangle \nonumber\\
    &=\langle(d-g^{T}\hat{\theta})^T T (d-g^{T}\hat{\theta})\rangle+\mu^T T\mu - 2\mu^T T\langle(d-g^{T}\hat{\theta})\rangle=T\Sigma+\mu^T T\mu
\end{align}

 From now on we reinsert the subscripts $A$ or $B$. 
Now we can consider  the  combination of $A$ and $B$, i.e.  $\beta=\Xi_{B}-\Xi_{A}+Y_{B}-Y_{A}$. 
We assume that the prior mean vector $\tilde\theta$ is the same for the common parameters, and zero for the additional B parameters. This is where the condition of nestedness enters and simplifies considerably the problem.
 Then clearly $u_A=u_B=u$.
 We have then
\begin{align}
(d_{m}-u_{m})T_{B,mk}(d_{k}-u_{k})-(d_{m}-u_{m})T_{A,mk}(d_{k}-u_{k}) = (d_{m}-u_{m})T_{mk}(d_{k}-u_{k})
\end{align}
where 
\begin{align}
T & =T_{B}-T_{A}=\Sigma^{-1}(g^{T}F^{-1}g|_{A}-g^{T}F^{-1}g|_{B})\Sigma^{-1}
\end{align}
In order to define the mean
\begin{equation}
\mu=\langle d-u\rangle=\langle d\rangle-u
\end{equation}
we need to specify whether we assume $A$ or $B$ to be true. Then
we have
\begin{equation}
\mu_{M}=g^T_{M}(\hat{\theta}_{M}-\tilde{\theta}_{M})
\end{equation}

  In the following, $\mu_M$ will take different values depending on whether one assumes $M=A$ or $M=B$ to be true. In the general case we have the following mean and variance:
\begin{align}
\langle\beta\rangle & =\langle\Xi_B\rangle-\langle\Xi_A\rangle+Y_{B}-Y_{A}=\mathrm{Tr}(T\Sigma)+\mu_M^{T}T\mu_M+Y_{B}-Y_{A}\label{eq:thmv}\\
V[\beta] & =\sigma_{\beta}^{2}=2\mathrm{Tr}(T\Sigma)^{2}+4\mu_M^{T}T\Sigma T\mu_M
\end{align}
Let us first consider the weak prior limit. 
In this case, $F\approx L$, and  $ T_M\Sigma \approx$ $\Sigma^{-1}(I-g_{M}^{T}L^{-1}g_{M}\Sigma^{-1})\Sigma= I-\Sigma^{-1}G_M=I-C_M$ \footnote{
One can also simplify the expression for the variance by noting that  $( T_M\Sigma)^2= T_M\Sigma$ and therefore $T_M\Sigma T_M= T_M\Sigma T_M \Sigma\Sigma^{-1}=(T_M\Sigma)^2\Sigma^{-1}=T_M\Sigma\Sigma^{-1} =T_M$.} 
 Since $C_M$ is a projector, its eigenvalues are 1 or 0.  Because of Eq. (\ref{eq:cproj}), we see that $C_M$ has $N_M$ unit eigenvalues and $N-N_M$ zero eigenvalues. Then the matrix
\begin{equation}
    -T\Sigma=-I+C_B+I-C_A=C_B-C_A
\end{equation}
has $r=N_B-N_A$ unit eigenvalues and the rest are zero. We conclude that $-\beta+(Y_B-Y_A)$ is a non-central $\chi^2$ variable with $\nu=r$ degrees of freedom and non-central parameter
\begin{equation}
    \lambda_M=-\mu_M^T T \mu_M
\end{equation}
 That is, $\beta$ is distributed as 
\begin{equation}
P(x;a,\nu,\lambda)=\frac{1}{2}e^{-(a-x+\lambda)/2}\left(\frac{a-x}{\lambda}\right)^{\nu/4-1/2}I_{\nu/2-1}(\sqrt{\lambda(a-x)})
\end{equation}
with $a=Y_{B}-Y_{A}$, and where $I$ is a modified Bessel function of the first kind. 
In the special case in which the elements of the prior mean vector $\tilde\theta$ coincide with model A's best fit parameters $\hat\theta_A$ for the common parameters, and vanishes for the extra B parameters, then $\mu_A=0$, so that $\lambda_A=0$, and $\beta$ is a standard $\chi^2$ variable up to shift and inversion, so its distribution is
\begin{equation}
P(x;a,\nu)=\frac{(a-x)^{\nu/2-1}e^{-(a-x)/2}}{2^{\nu/2}\Gamma(\nu/2)}
\end{equation}

 Finally, if the prior is not negligible, then the eigenvalues of $-T\Sigma$ are no longer 1 or 0's, and $\beta$ can be written as a sum of non-central $\chi^2$ variables. This sum follows a generalized $\chi^2$ distribution, as we discuss next.

We need to find the  distribution of a general quadratic form
\begin{equation}
Q=(d_{m}-u_{m})T_{mk}(d_{k}-u_{k})=X^{T}TX
\end{equation}
This can be obtained as follows (see \cite{10.1093/biomet/48.3-4.419,DUCHESNE2010858}). The mean of $X$ is $\mu=\langle X\rangle=g^{T}(\hat{\theta}-\tilde{\theta})=f-u$.
The covariance matrix of $X$ is
\begin{align}
\Sigma & =\langle(d-u-\mu)(d-u-\mu)^{T}\rangle\\
 & =\langle(d-f)(d-f)^{T}\rangle
\end{align}
Let now $C$
be a matrix such that (Cholesky decomposition)
\begin{equation}
CC^{T}=\Sigma
\end{equation}
and $P$ an orthonormal matrix (i.e., $PP^{T}=I)$ that diagonalizes
$CTC^{T}$, i.e. 
\begin{equation}
PCTC^{T}P^{T}=D=\mathrm{diag}(\alpha_{1},\alpha_{2}..\alpha_{p})
\end{equation}
where the eigenvalues of $CTC^{T}$ $\alpha_{i}$ are non-zero for
$i=1...r$ and zero for $i=r,..,p$. The number of non-zero eigenvalues should
be $r=N_{B}-N_{A}$. Let also define the two vectors $Y=P(C^{T})^{-1} X$ and $\xi=P(C^{T})^{-1}\mu$.
Then one can see that $Y$ is a vector of Gaussian variables with  mean vector $\xi$ and covariance matrix $\Sigma_Y=I$. Therefore $Q$ can be written as a linear combination of non-central, uncorrelated $\chi^2$ variables,
\begin{equation}\label{eq:lincombnoncen}
Q=X^{T}TX=\sum_i^r\alpha_i Y_i^2=\sum_{i}^{r}\alpha_{i}\chi_{1}^{2}(\lambda_{i})
\end{equation}
where $\chi_{1}^{2}(\lambda)$ are non-central $\chi^{2}$ variables
with one degrees of freedom and non-central parameter 
\begin{equation}
\lambda_{i}=\xi_{i}^{2}
\end{equation}
Then, in general, $Q$ is distributed as a generalized $\chi^2$ variable.  The  characteristic function (CF) for the generalized $\chi^{2}$ variable can be obtained analytically  \cite{10.1093/biomet/48.3-4.419}:
\begin{equation}
\phi(t)=\frac{\exp\left\{ it\sum_{i}^{r}\frac{\lambda_{i}\alpha_{i}}{1-2i\alpha_{i}t}\right\} }{\prod_{i}^{r}(1-2i\alpha_{i}t)^{\frac{1}{2}}}
\end{equation}
However, one cannot derive analytically neither  the probability density function (PDF) nor the cumulative distribution function. One can numerically determine the PDF performing an inverse Fourier transform or utilizing existing  codes (e.g. \cite{DUCHESNE2010858}).

We have shown earlier  that in the limit of weak prior, the eigenvalues are indeed unity. In this case,  $Q$ is a non-central variable with $\lambda=\xi^T \xi$ and $r$ dofs. Finally, if the eigenvalues are unity and $\lambda=0$, $Q$ is a standard $\chi^2$ variable with $r$ dofs. This shows how the generalized $\chi^2$ distribution reduces to one of the simpler forms. 

In the case $r=1$, $Q$ is  a single non-central variable
with one dof and 
\begin{equation}
\lambda=\xi^{2}
\end{equation}
In this case, the distribution of $\beta=\alpha\chi_{1}^{2}(\lambda)+a$
is obtained from the non-central distribution $P(\chi_{1}^{2}(\lambda))$
as 
\begin{equation}
P_\beta(\beta )=P(\chi_{1}^{2}(\lambda))|\frac{d\chi_{1}^{2}(\lambda)}{d\beta}|=\frac{1}{|\alpha|}P(\frac{\beta-a}{\alpha})
\end{equation}

 To summarize, if one assumes a Gaussian prior and Gaussian data with mean model linear in the parameters, then the logarithm of the Bayes factor is a random variable distributed like one of three kinds of $\chi^2$ variables (central, noncentral, or generalized $\chi^2$), depending on the mean and width of the prior relative to the likelihood.
 \footnote{Clearly, the logarithm of the evidence is also a $\chi^2$ variable, now with $N-N_M$ dofs. One could also envisage a frequentist test based on the evidence of a single model, without comparing it to an alternative. In this paper, however, we explore only the properties of Bayes' ratio.}

 The Bayes' distributions can now be employed to build our FB method, i.e. to test the hypothesis that, given that model M is true, the data are consistent with M rather than with an alternative M'. 
 Since a $p$-value test can only reject hypotheses, and not assess whether they are true, we need to make two tests when comparing A versus B. In the first, we assume A to be true, so that the null hypothesis is $H_0:$ (A is true), and we either reject or non-reject A \footnote{In this paper, we will adopt the terminology commonly used by in cosmological statistical data analysis for frequentist hypothesis testing, replacing the terms Type I/II error with the expressions like "rejecting" or "not rejecting the null hypothesis" $H_{0}$.}; in the second test, we assume B to be true, so $H_0:$ (B is true), and again we can reject or non-reject B. Together, the two tests can produce four distinct outcomes: reject both models, reject neither of them, reject only A or only B.
 One can consider both one-tailed tests, as commonly done for likelihood ratio tests, or two-tailed tests, as is typically the case for $\chi^2$ tests. In this second case, the point of view is that too good fits are rejected because incompatible with the assumed data uncertainties.
 Here, however, we only discuss one-tailed tests for simplicity.
 The entire procedure will be illustrated below in Sec. \ref{sec:pol} and \ref{sec:sn}.

\section{Fisher approximation}
\label{sec:Fisher}

In many realistic applications, parameters do not enter linearly in the likelihood. 
The fitting function $f$ can however  be linearized near the best fit as
\begin{equation}\label{eq:fiapprox}
f(x,\theta)\approx f(x)|_{\theta=\hat\theta}+\sum_\alpha \frac{\partial f}{\partial \theta_\alpha}|_{\theta_\alpha=\hat\theta_\alpha}(\theta_\alpha-\hat\theta_\alpha)
\end{equation}
The Fisher matrix for Gaussian data with parameters only in the mean vector $f_i$ is then
\begin{equation}
    F_{\alpha\beta}=\left(\frac{\partial f_i}{\partial\theta_\alpha}\Sigma^{-1}_{ij}\frac{\partial f_j}{\partial\theta_\beta}\right)_{\theta_\alpha=\hat\theta_\alpha}
\end{equation}
Then we see that we can identify
\begin{equation}
    g_{\alpha i}=\frac{\partial f(x_i)}{\partial \theta_\alpha}|_{\theta_\alpha=\hat\theta_\alpha}
\end{equation}
We can then use a Fisher matrix approximation to obtain the matrix $L$:
\begin{equation}
    L_{\alpha\beta}=g_{\alpha i}\Sigma^{-1}_{ij}g_{\beta j}
\end{equation}
and the problem is reduced to the one of the previous sections. Whether this approximation is acceptable, however, should be analysed on a case-by-case basis. We apply this Fisher approximation in Sec. \ref{sec:sn}.

\section{A frequentist model-comparison test}
The use of the  $\beta$ distribution to see whether models $A$ or $B$ are compatible with the data is an absolute test of compatibility of data with models (goodness of fit). But we can also directly compare $A$ with $B$ and see which one is to be considered a better fit to the data (model comparison). The result of this model comparison will be a  statement about which model better explains the data, even if both are, on absolute terms, unacceptable fits. These two tools, together, form the content of our FB method.

 In Bayesian context,  $b=e^{\beta/2}$ is not a random variable. Its
value expresses the odds ratio of $A$ versus $B$, i.e. the ratio
of the probabilities that, given the observed data $D$, the model
$A$ or the model $B$ is to be preferred,
\begin{equation}
b (D)=e^{\beta/2}=\frac{P(A)}{P(B)}
\end{equation}
The value of $\beta$ is then compared to an empirical scale (the Jeffreys scale or one of its variants) to convert the number into an expression of ordinary language. The conditioning over $D$ will be implicit in the following.

 Alternatively, we can also  convert $b$ into the probability of $A$ given that the
possible models are only $A$ or $B$ as follows (the subscript $b$ here
stands for Bayesian):
\begin{equation}
R_{b}(A)\equiv P(A|A\mathtt{or}B)=\frac{P(A\mathtt{and}(A\mathtt{or}B))}{P(A\mathtt{or}B)}=\frac{P(A)}{P(A)+P(B)}=\frac{b}{1+b}=(1+e^{-\beta/2})^{-1}
\end{equation}
In this way, Bayes' ratio can be interpreted as the probability of
``event'' $A$ within a set of possible disjoint events composed only by
$A$ and $B$. This probability over the space of events is well normalized:
in fact, one also has
\begin{equation}
R_{b}(B)\equiv P(B|A\mathtt{or}B)=\frac{P(B)}{P(A)+P(B)}=1-R_{b}(A)
\end{equation}
Following this point of view, the Jeffreys scale can be replaced by ordinary
probability thresholds. For instance we can accept model $A$ versus
$B$ at 68\% c.l. if
\begin{equation}
R_{b}(A)=\frac{b}{1+b}\ge0.68
\end{equation}
or $b\ge2.125$ (``barely worth mentioning'' in Jeffreys' language),
at $95\%$ c.l. if $b\ge19$ (``strong'') and at $99.7\%$ if $b\ge332$
(``decisive''). Naturally, if there are other models to consider, and not just $A$ and $B$ (as is almost always the case), then this scale is hardly meaningful because it depends on the arbitrary number of models we decide to include. 
The main problem, however, as we already emphasized, is that $R_b$ depends on the prior also in the limit of weak priors:
\begin{equation}
    R_b=(1+e^{-\beta/2})^{-1}=(1+e^{-(\chi^2_B-\chi^2_A+Y_B-Y_A)/2})^{-1}=(1+\sqrt{\frac{|F_{A}||P_{B}|}{|P_{A}||F_{B}|}}e^{-\frac{1}{2}(\chi_{B}^{2}-\chi_{A}^{2})})^{-1}
\end{equation}
For instance, let us assume that  model $B$ is identical to model $A$ except for an additional parameter that has an uncertainty $\sigma_B$ 
and a
prior with a very large uncertainty $\sigma_{P,B}\to\infty$. Then, if both models have a diagonal Fisher matrix, we can write $F_B=F_A\sigma_B^{-2}$ and $P_B=P_A\sigma_{P,B}^{-2}$, and one gets
\begin{equation}
    R_b(A)\approx 1-(\frac{\sigma_B}{\sigma_{P,B}})^2 e^{-\frac{1}{2}(\chi^2_B-\chi^2_A)}\to 1
\end{equation}
This shows that model $A$ is always favoured in this limit, regardless of its quality. That is, the addition of a parameter with a very weak prior, maybe because $B$ is a new model with little experimental history, is  discarded a priori. This carries the risk of artificially enhancing trust on established models when confronted with unbeaten paths (see App. \ref{sec:simple} for an example).

 In the frequentist interpretation, the Bayesian odds ratio should
be replaced by the ratio of the distributions of $\beta$ when assuming
either $A$ or $B$. We can then form the frequentist (subscript $f$) analog of $R_{b}$
as
\begin{equation} \label{eq:frequentist_odds_ratio}
R_{f}(A)=\frac{P_{A}(\alpha_A\beta+a)}{P_{A}(\alpha_A \beta+a)+P_{B}(\alpha_B\beta+a)}
\end{equation}
and of course its complement, $R_{f}(B)=1-R_{f}(A)$. 
 We denoted with
$P(\alpha_M\beta+a)$ a   distribution ($\chi^2$ or otherwise) for
the variable $\chi^{2}=\alpha_M\beta+a$ with $M=A,B$. 
Since in the weak prior limit $\alpha_M\beta+a=\chi^2_A-\chi^2_B$, we see that $R_f$, contrary to $R_b$, becomes independent of the prior.
A value of $R_f(A)$ larger than $1/2$ means that $A$ is favoured with respect to $B$; viceversa, $R_f(A)< 1/2$ means that $B$ is favoured. However, here we encounter the same caveat as in the Bayesian case, namely, that $R_f$ can be directly compared to a Gaussian scale  only if the models $A$ and $B$ are exhaustive of the model landscape.

\section{Applications: polynomial fits}
\label{sec:pol}

Now that we know the distribution of the statistic $\beta$, we can
form an analytical $p$-value test.   Let  us assume as null hypothesis $H_{0}:$ model $A$
is better than B in explaining the data, and as alternative hypothesis $H_{1}$: model $B$ is better than $A$.
We can then find a specific value of $\beta$ from a given dataset and  compare it to the $\beta$ distribution.

\begin{figure}
\begin{centering}
\includegraphics[width=16cm]{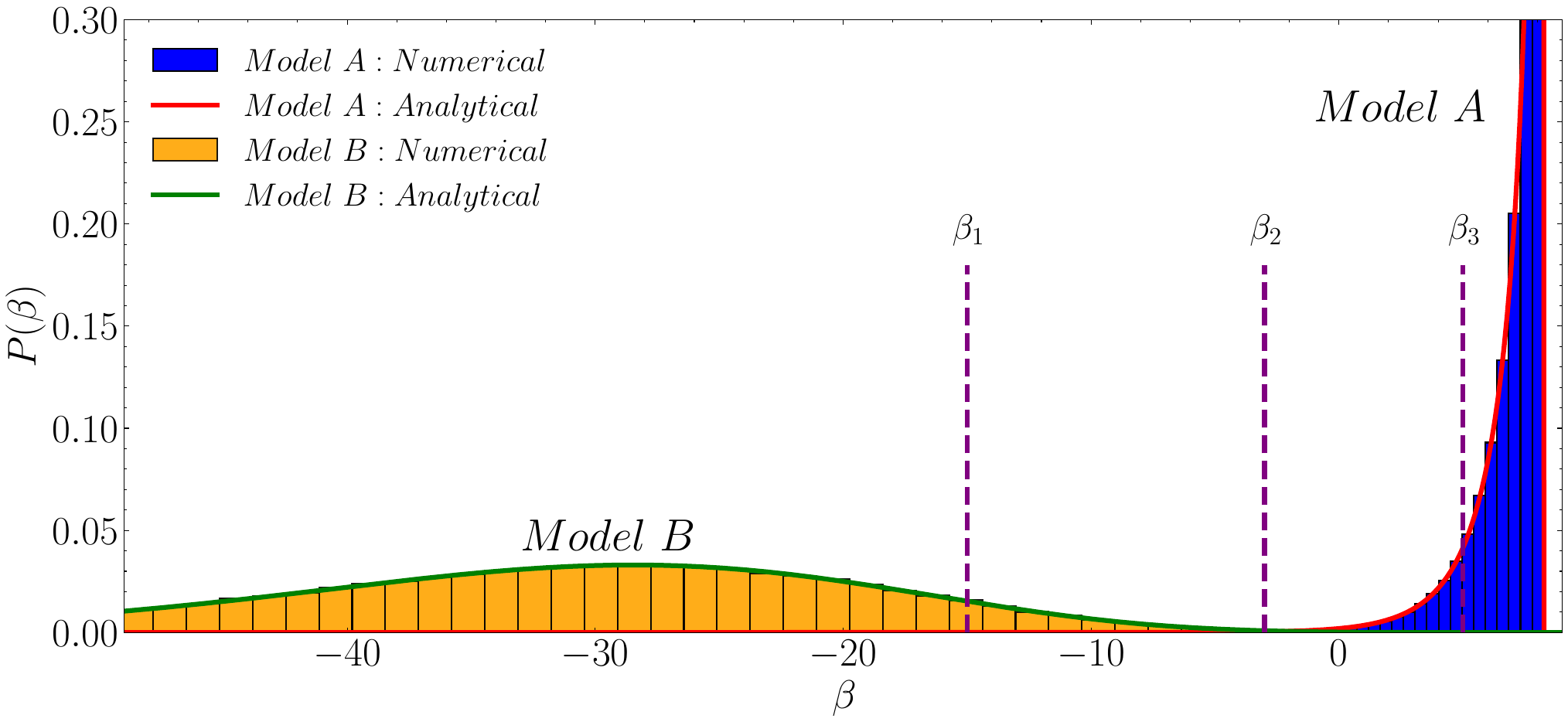}
\par\end{centering}
\caption{\label{fig:Distribution-of-}
Distribution of $\beta$ assuming model
$A$, a straight line  with two parameters (right) or model $B$, a quadratic polynomial with three parameters
(left), with $c=1$. The prior has been taken to be very broad. The histograms
have been obtained by a numerical simulation. Here and in the subsequent similar plots, the red and green curves are the $\chi^2$ analytical results.  If an experiment finds a value $\beta_{1}$,
$A$ is rejected; if it finds $\beta_{2}$, both $A$ and $B$ are
rejected; if it finds $\beta_{3}$,
$B$ is rejected. }

\end{figure}

\begin{figure}
\begin{centering}
\includegraphics[width=16cm]{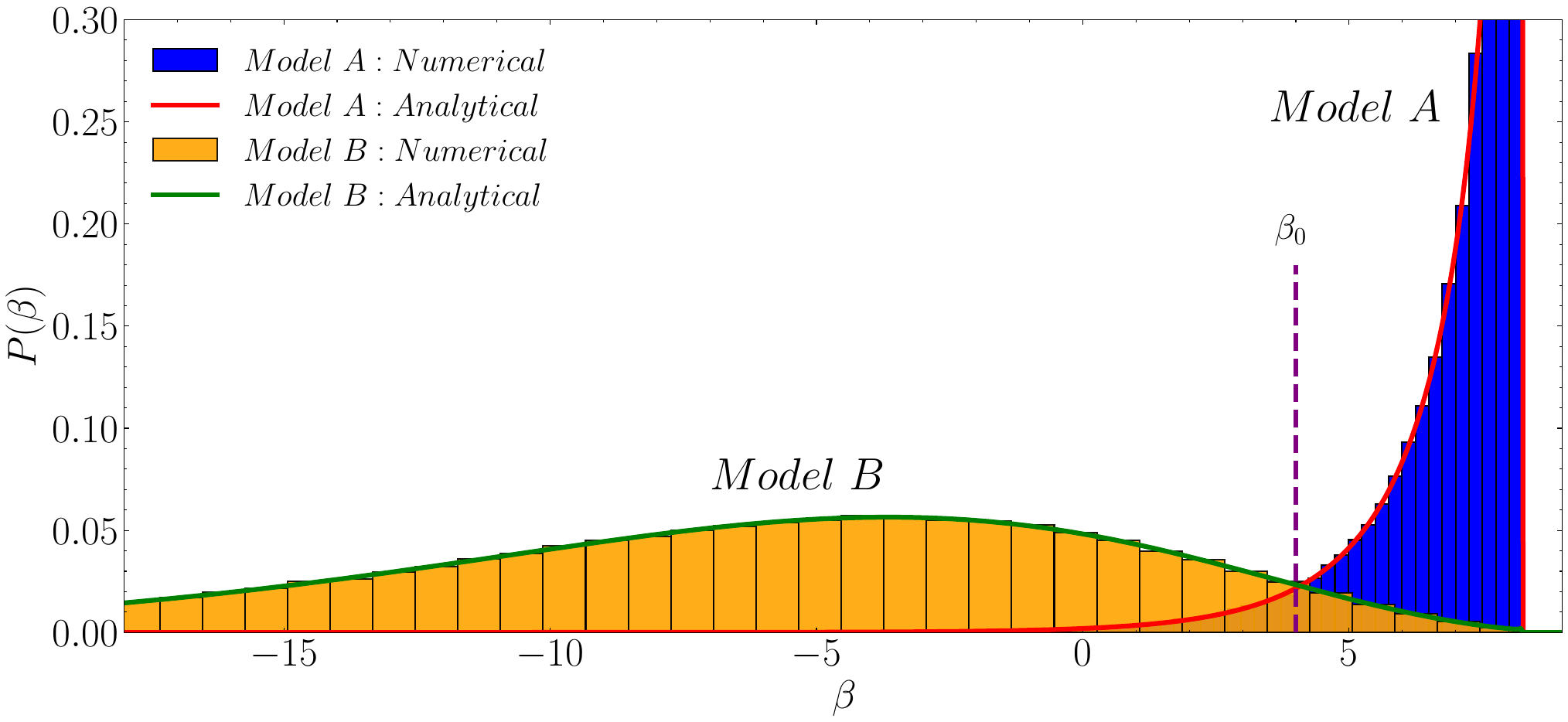}
\par\end{centering}
\caption{\label{fig:Distribution-of--1}Same as the previous plot, but for two models that are closer ($c=0.6$).  If an experiment finds a value $\beta$ as marked by the  dashed vertical
line, neither model can be rejected to a high significance.}
\end{figure}

 As a first illustration of the method,
we define two models: model A (a straight line $f=a+bx$) and model B (a quadratic function $f=a+bx+cx^2$, with the same $a,b$ as for model A). We take as prior for $a,b,c$ a Gaussian with mean $a=1,b=1$ and two values of $c$, and diagonal prior covariance with a very broad variance, $\sigma_P= 10$ for each parameter (to be compared to uncertainties between 0.1 and 1 for the three parameters).
Then we generate many sets of  Gaussian data points distributed around model A  with a given data covariance matrix $\Sigma$ (we put for simplicity $\Sigma=I$, but we tested also correlated cases). 
Every time a data set is generated, the best fit parameters $\hat a,\hat b,\hat c$ are obtained by fitting first with a straight line and then with a quadratic function, and the resulting $\beta$ is calculated. This gives the distribution of $\beta$ assuming A is correct. Then we repeat generating many data sets around model B (with the same prior and data covariance matrix), and obtain the distribution of $\beta$ when B is assumed correct. 
The parameter $c$ is a measure of how different  the two models are, and therefore of how much the two distributions overlap.
Putting as means $a=b=1$,  we find the distributions as in Figs. \ref{fig:Distribution-of-}  (with $c=1$) and \ref{fig:Distribution-of--1} (with $c=0.6$). 

 We now assume that the data give a particular value of $\beta$, say $\beta=\beta_i$.
As illustrated in the figures, there are four possibilities: A is rejected ($\beta_1$), B is rejected ($\beta_3$), both models are rejected ($\beta_2$), neither model is rejected ($\beta_0$ in Fig. \ref{fig:Distribution-of--1}). In this last case, Fig. \ref{fig:Distribution-of--1}, if the data give $\beta=0$, i.e. equal evidences for the two models, the Bayesian test would be inconclusive, while here we see that A is rejected but B is not.
In terms of $p$-values,  we obtain $p\sim 10^{-7}$ for $\beta_1$ assuming A is true; we obtain $p\sim10^{-6}$ for $\beta_3$ assuming B is true; and we obtain $p=0.00063 \,\,(0.0017)$ for $\beta_2$ assuming A (respectively B).
Notice that the $p$-values are obtained integrating the distribution in the interval $(-\infty,\beta_i)$ when assuming A, and in the interval $(\beta_i,\infty)$ when assuming B.

\begin{figure}
\begin{centering}
\includegraphics[width=10cm]{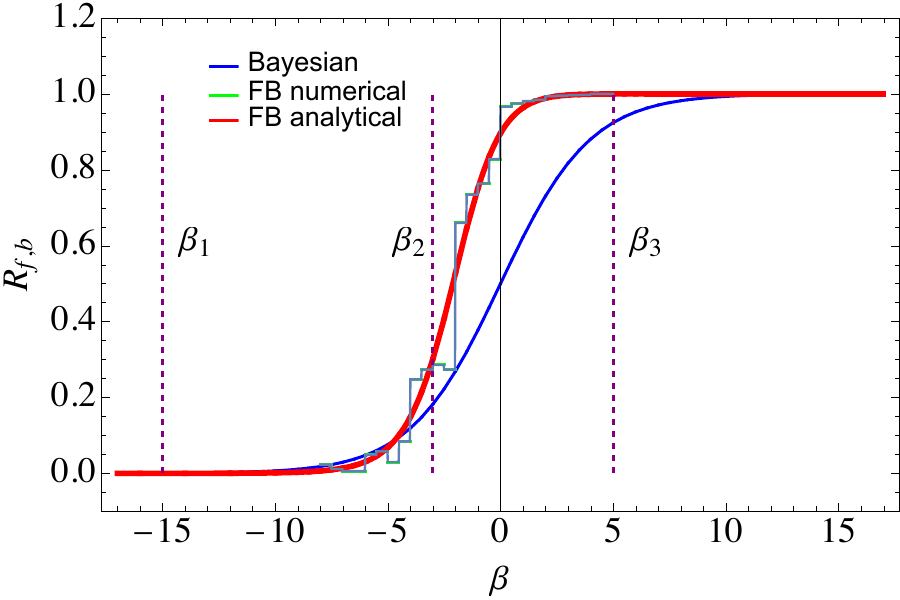}
\par\end{centering}
\caption{\label{fig:rf-1} For the same settings as in Fig. \ref{fig:Distribution-of-}, here we plot 
$R_{f}$ (theoretical distribution: red line; numerical distribution:
green line) and $R_{b}$ (blue line)  versus $\beta$. }
\end{figure}

\begin{figure}
\begin{centering}
\includegraphics[width=10cm]{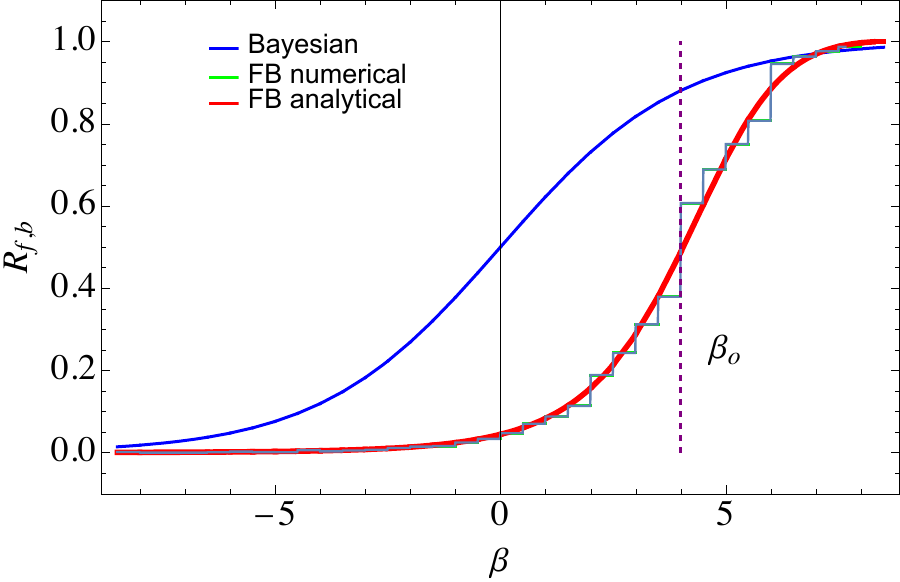}
\par\end{centering}
\caption{\label{fig:rf-06}For the same settings as in Fig. \ref{fig:Distribution-of--1}, here
$R_{f}$ (theoretical distribution: red line; numerical distribution:
green line) and $R_{b}$ (blue line) are shown versus $\beta$.}
\end{figure}

 In Figs. \ref{fig:rf-1} and \ref{fig:rf-06} we illustrate the model comparison test for the same settings as the previous figures. Clearly, $\beta_1$ favors $B$, $\beta_3$ favors $A$, while $\beta_0,\beta_2$ are non-committal. If we had obtained the value $\beta=0$ for the case of Fig. \ref{fig:rf-1}, then the Bayesian model comparison would have been inconclusive, while in our FB test it would have been quite in favor of $B$ because essentially incompatible with the data, as can be seen in Fig. (\ref{fig:Distribution-of--1}).

Plots (\ref{fig:Distribution-of-}-\ref{fig:rf-06}) represent therefore a full analysis of the models. The $p$-value test gives the absolute quality of the fits, while the probability ratio $R_f$ expresses the relative advantage of a model over another one, regardless of whether they are a good fit or not.

\begin{figure}
\begin{centering}
\includegraphics[width=16cm]{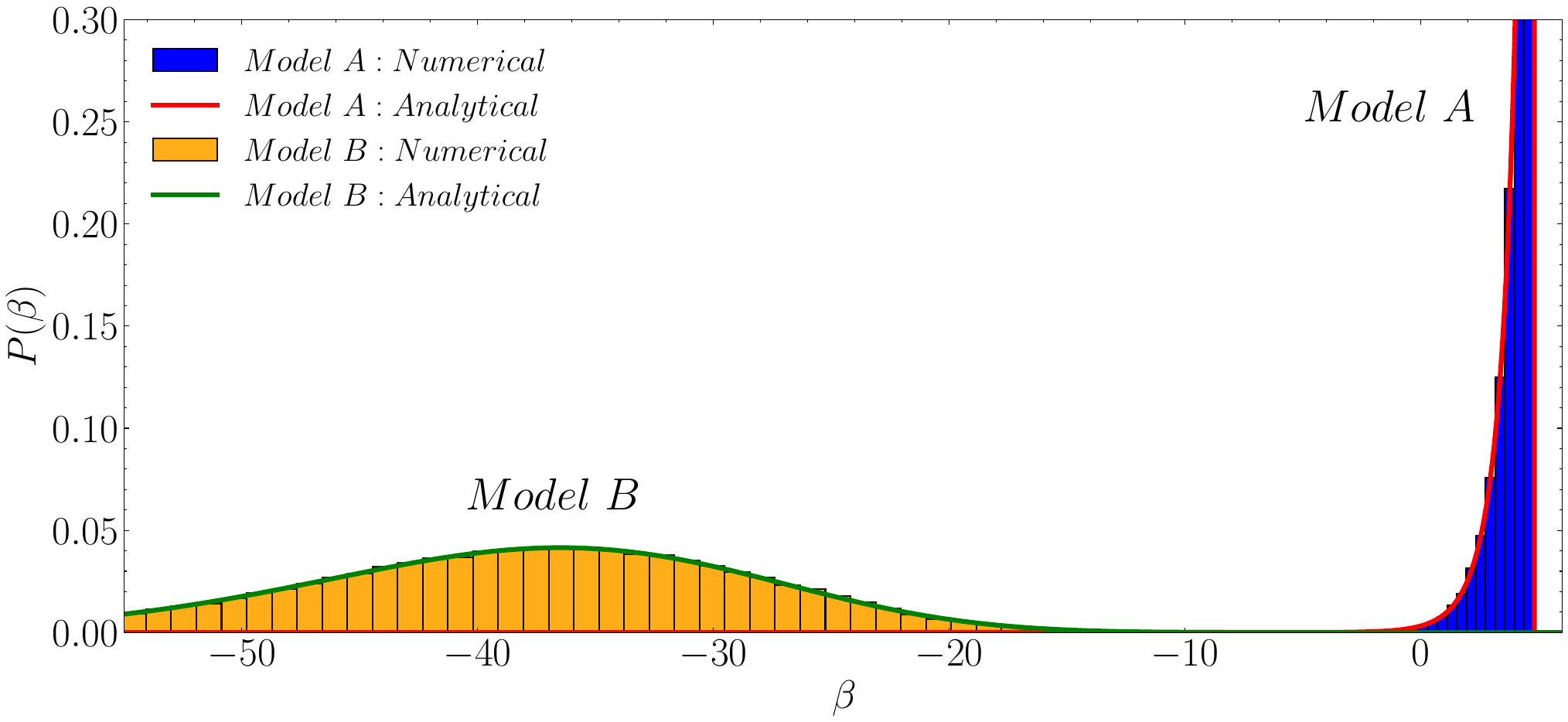}
\par\end{centering}
\caption{\label{fig:strong} Same as Fig. \ref{fig:Distribution-of--1}, but for a strong prior $\sigma_P=1$.  }
\end{figure}

\begin{figure}
\begin{centering}
\includegraphics[width=16cm]{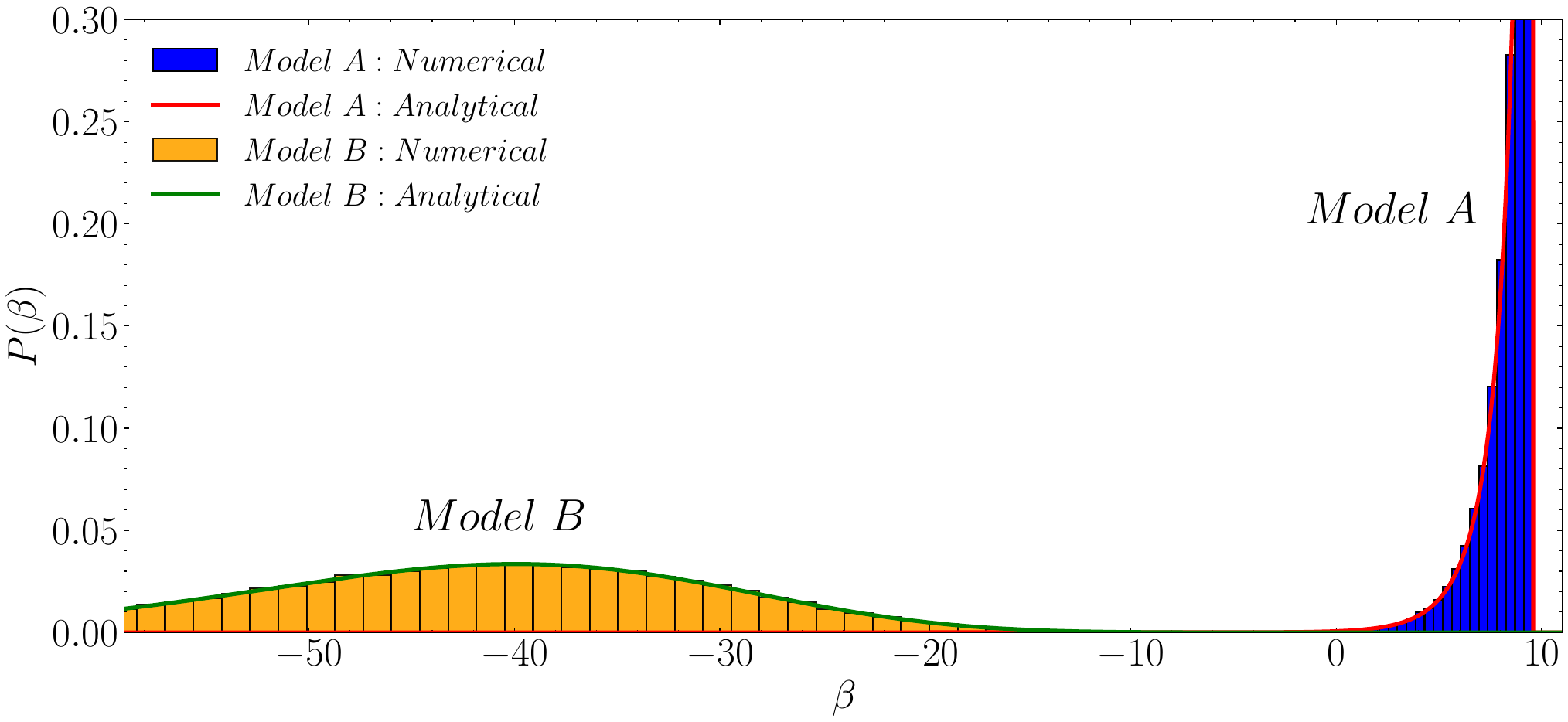}
\par\end{centering}
\caption{\label{fig:cubic-comp} Distribution of $\beta$ with modified model B, a cubic function, with $c=0.04$ and $d=0.06$. The prior used here is $\sigma_{P}=0.5$. In this case, the distribution is a generalized $\chi^2$.}
\end{figure}

Further, in Fig. \ref{fig:strong} we show how the non-central $\chi^2$  distribution  reproduces the numerical distribution in the same case as Fig.  \ref{fig:Distribution-of--1} but now with a tight prior ($\sigma_P=1$). Since the prior is strong and centered on model A parameters, the range  of $\beta$ that favors model B is now much reduced with respect to  Fig. \ref{fig:Distribution-of--1}. 
Finally, in Fig. \ref{fig:cubic-comp}, we modified model B to a cubic function ($f=a+bx+cx^{2}+dx^{3}$), and the corresponding numerical distribution was generated as described earlier. In this case, however, the analytical distribution takes the shape of a generalized $\chi^{2}$.


\section{Applications: supernovae}
\label{sec:sn}

 In this section, we apply the FB approach to real supernovae data. We have used supernovae data from the Union2.0 catalogue \cite{AmanullahLidman2010}, which consists of a total of 557 supernovae observations. The models we are comparing are $\Lambda$CDM and $w$CDM.

 The absolute magnitude of a supernova can be related to the luminosity distance using the equation:
\begin{equation}
\mu=5\log_{10}(d_{L}) + 25=m-M
\end{equation}
Here, apparent magnitudes ($M$) and absolute magnitudes ($m$) are the observables for a supernova, and the luminosity distance $d_{L}$ is measured in Mpc.
The luminosity distance $d_{L}$ can be given as (we assume flat space and adopt  units such that $c=1$):
\begin{equation}\label{eq:LCDMdist}
d_{L}(z)=\frac{(1+z)}{H_{0}}\int_{0}^{z}\frac{dz^{\prime}}{E(z')},
\end{equation}
where $E^2(z)=\Omega_{m}(1+z)^{3}+\Omega_{\Lambda}$  for $\Lambda$CDM and  $E^2(z)=\Omega_{m}(1+z)^{3}+\Omega_{de}(1+z)^{3(1+w)}$  for $w$CDM
and where $H_{0}$ is the Hubble constant,
 $w$ the dark energy equation of state, and 
 $\Omega_m, \Omega_{\Lambda},\Omega_{de}$ the present matter, cosmological constant, and dark energy  density parameters, respectively.

\begin{figure}
\begin{centering}
\includegraphics[width=15cm]{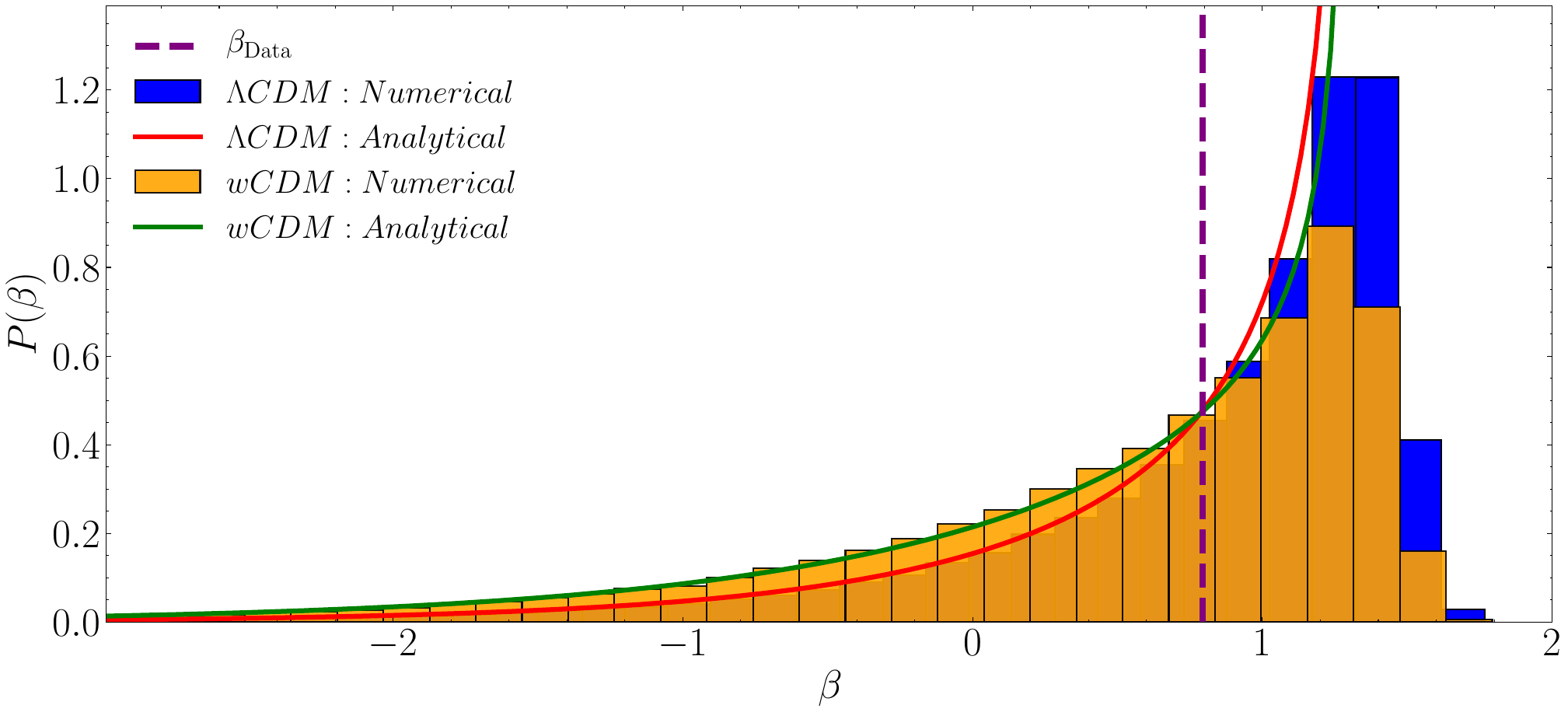}
\par\end{centering}
\caption{\label{fig:betaSN} Distribution of $\beta$ for supernovae Ia.  For this simulation, a strong prior configuration was employed with $\sigma_{\Omega_{m}}^{2}=0.005$  and $\sigma_{w}^{2}=0.08$.
}
\end{figure}

\begin{figure}
\begin{centering}
\includegraphics[width=15cm]{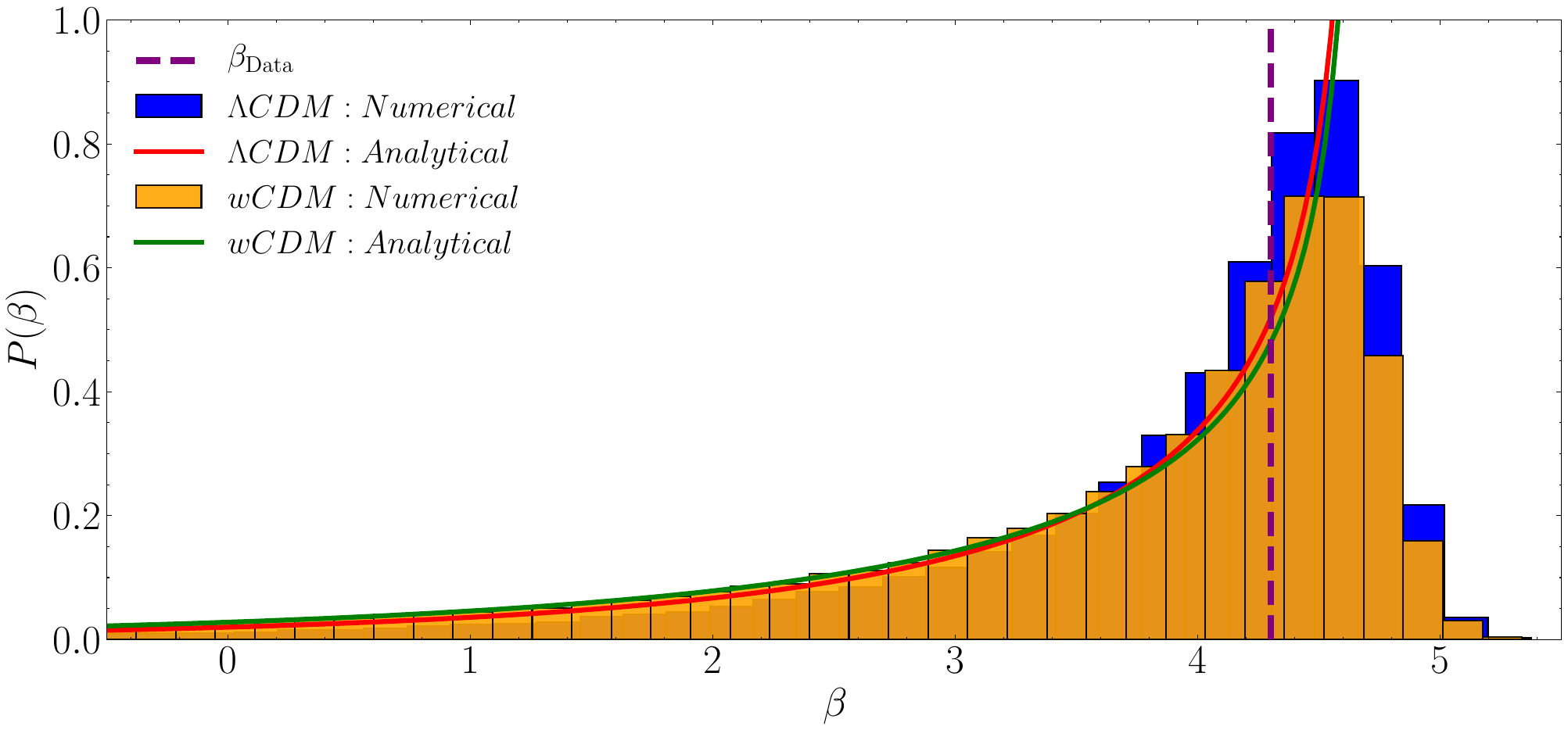}
\par\end{centering}
\caption{\label{fig:betaSN_strong}Similar to  Fig. \ref{fig:betaSN}, but for  weak priors:  $\sigma_{\Omega_{m}}^{2}=0.5$ and $\sigma_{w}^{2}=8$.}
\end{figure}

 Now, we can formulate the likelihood Eq. (\ref{eq:genlik}) for our analysis as:
\begin{equation}
    \mathcal{L}=M\exp[-\frac{1}{2}(m_{\rm data}-5\log_{10}d_{L}^{\ast}+\alpha)_{i}\Sigma_{ij}^{-1}(m_{\rm data}-5\log_{10}d_{L}^{\ast}+\alpha)_{j}]
\end{equation}
where $d^*_L=H_0 d_L$, the indexes $i,j$ run over the data points, and $\alpha=5\log_{10}H_{0}+25-M$. We can marginalize over $\alpha$ 
and the new likelihood can be written as:
\begin{equation}\label{eq:marglik}
    \mathcal{L}=M^{\ast}\exp[-\frac{1}{2}(m_{\rm data}-5\log_{10}d_{L}^{\ast})_{i}K_{ij}^{-1}(m_{\rm data}-5\log_{10}d_{L}^{\ast})_{j}]
\end{equation}
where
\begin{equation}
K_{ij}^{-1}\equiv \Sigma_{ij}^{-1}-\frac{u_{p}\Sigma_{pi}^{-1}u_{n}\Sigma_{nj}^{-1}}{S_{0}}
\end{equation}
with
\begin{align}
u_{i}&=\{1,1,1...\}\\
S_{0} & =u_i \Sigma_{ij}^{-1}u_j=\sum_{ij}\Sigma_{ij}^{-1}
\end{align}
Note that the matrix $K^{-1}$ is singular, so one cannot invert it to obtain $K$. This causes no problem, however, since we only need $K^{-1}$. 
Bayes' ratio was computed for the data using this likelihood and two different priors:  a  strong prior case, in which the variances were set at $\sigma_{\Omega_{m}}^2=0.005$ and $\sigma_{w}^2=0.08$, and a  weak prior case, with $\sigma_{\Omega_{m}}^2=0.5$ and $\sigma_{w}^2=8$. We obtained a value of $\beta_{\rm data}=0.79$ for strong priors and $\beta_{\rm data}=4.30$ for weak priors  (see Figs. \ref{fig:betaSN} and \ref{fig:betaSN_strong}). In standard Bayesian analysis, a $\beta_{\rm data}$ value of $0.79$ for strong priors would be considered ``barely worth mentioning" evidence in favour of $\Lambda$CDM, while a value of $4.30$ for weak priors would be deemed ``substantial" evidence. Our method aligns with the standard test in the case of strong priors. However, we contend that in the case of weak priors, both models should be regarded on a similar footing, contrary to what the standard test suggests.

For the $\beta$ distribution, first, a model was supposed to be true; let us say $\Lambda$CDM. Subsequently, a synthetic dataset was created by selecting random $\mu$ values from a multivariate Gaussian distribution assuming $\Lambda$CDM. The mean of this distribution was determined using Eq. (\ref{eq:LCDMdist}), and the covariance matrix was taken from the Union2.0 catalogue. From Eq. (\ref{eq:marglik}), a likelihood and, consequently,  $\beta$ are computed for this simulated dataset. This procedure was repeated 100000 times, resulting in a distribution of $\beta$ values denoted as $P(\beta)$. A similar procedure is then replicated, assuming the $w$CDM model to be true. 

 Our analytical approach works for linear models, but the parameters enter the models $\Lambda$CDM
and $w$CDM in a non-linear way. Linearization of our models
can be done by approximating them by first-order Taylor expansion
at the parameter best-fit values Eq. (\ref{eq:fiapprox}), 
\begin{equation}\label{eq:dlapprox}
    d_{L}^{*}(z;\theta_{i})\approx d_{L}^*(z)|_{\theta=\hat{\theta}}+\sum_{i}\frac{\partial d_{L}^*}{\partial\theta_{i}}|_{\theta_{i}=\hat{\theta_{i}}}(\theta_{i}-\hat{\theta_{i}})
\end{equation}
where $\theta_{i}$ are $(\Omega_{m})$ and $(\Omega_{m},w)$ for $\Lambda$CDM
and $w$CDM, respectively.  This is a good approximation:  the relative error between $d_{L, \rm exact}^\ast$ (Eq.\ref{eq:LCDMdist}) and $d_{L, \rm approx}^\ast$ (Eq.\ref{eq:dlapprox}) for both $\Lambda$CDM and $w$CDM models is less $1\%$ for $z\leq2.5$.

In this scenario, where we have two linear models with the number of parameters $N_{\Lambda\rm CDM}=1$ and $N_{w\rm CDM}=2$, the analytical
approximation of the $\beta$ distribution can be represented by a non-central
$\chi^{2}$ distribution with $\nu=N_{w\rm CDM}-N_{\Lambda\rm CDM}=1$ degrees
of freedom and non-centrality parameter $\lambda$. This analytical distribution is shown in Figs. \ref{fig:betaSN} and \ref{fig:betaSN_strong}.

\begin{figure}
\begin{centering}
\includegraphics[width=15cm]{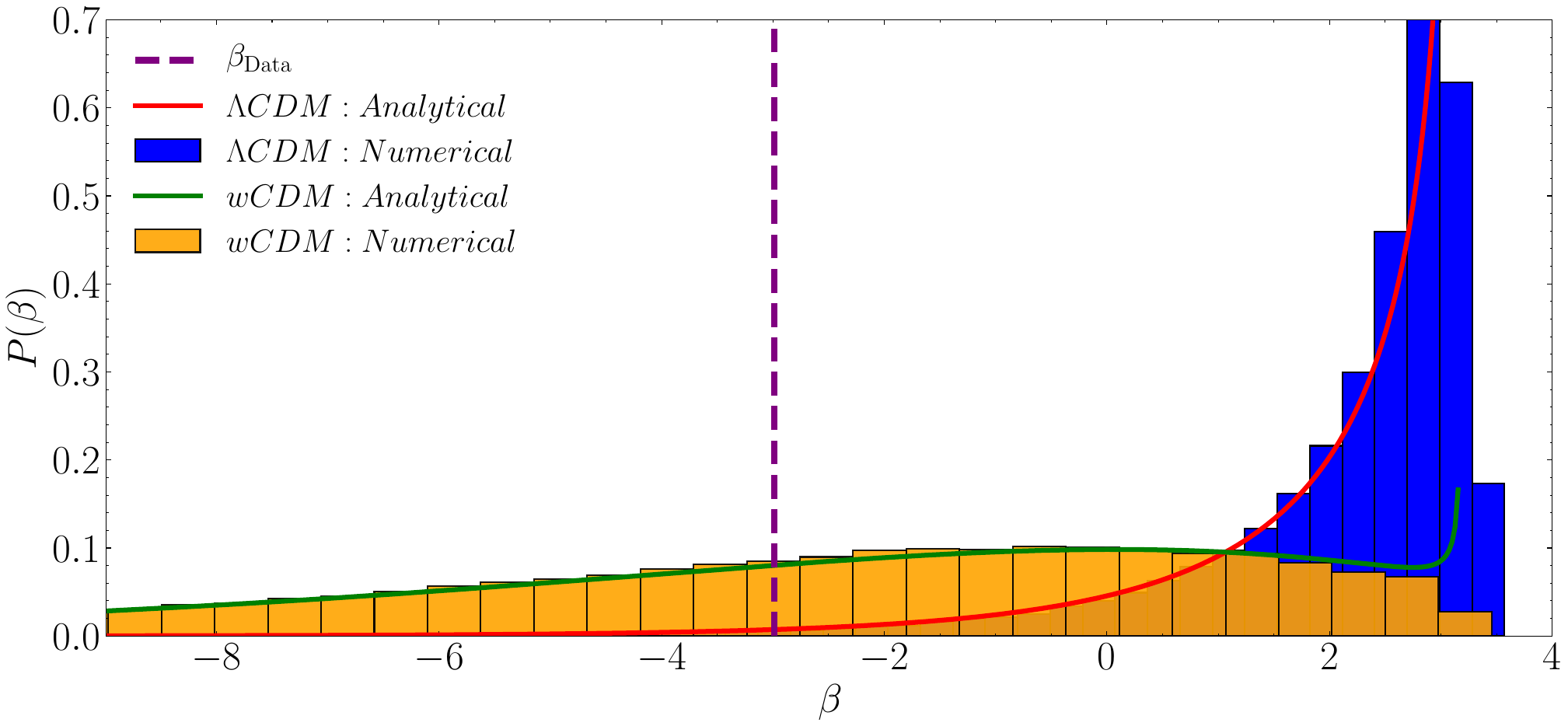}
\par\end{centering}
\caption{\label{fig:betaSN_future} Similar to  Fig. \ref{fig:betaSN}, but now with the supernovae covariance reduced by a factor of 20 (i.e., $C_{ij}/20)$. The two distributions are now quite different, and it would be possible to clearly differentiate the two competing models.  }
\end{figure}

\begin{figure}
\begin{centering}
\includegraphics[width=15cm]{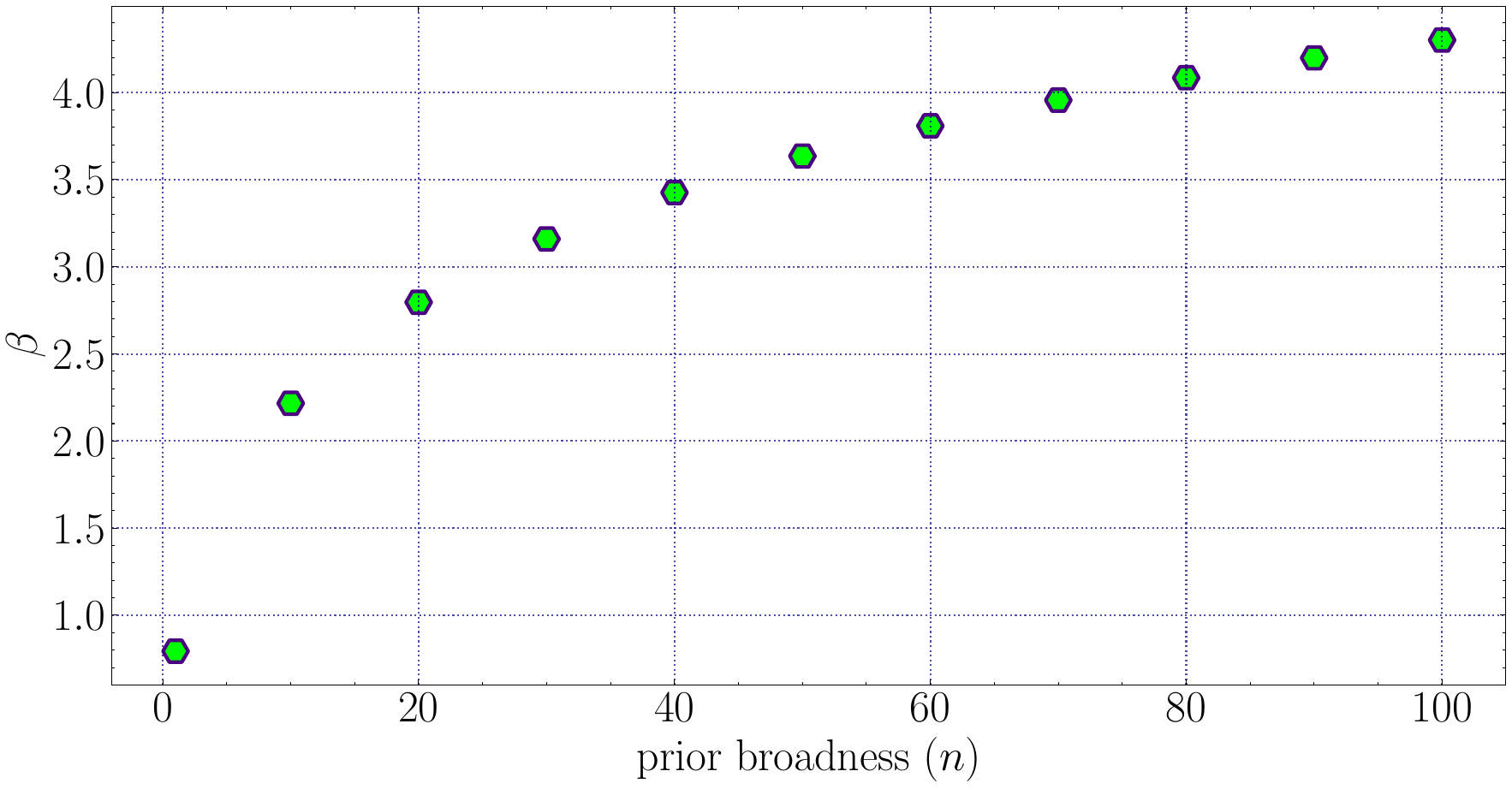}
\par\end{centering}
\caption{\label{fig:pridep1} Dependence of $\beta$  on different priors. The Gaussian variance of priors for each $\beta$ is taken as $\sigma_{\Omega_{m}}^2 = n \times 0.005$ and $\sigma_{w}^2 = n \times 0.08$, where $n =$ prior broadness.}
\end{figure}

The best-fit value of the additional parameter in the $w$CDM model ($\hat{w}=-1.15$) is very close to the value for the $\Lambda$CDM model ($w=-1$), making both distributions very similar to each other. This close alignment between the distributions makes it challenging to conclusively comment on the accuracy of either cosmological model at present. However, we anticipate that future supernovae observations, with reduced observational uncertainties, will enable a more distinct separation between two distributions.  As a purely illustrative example, if the SN magnitude uncertainties are decreased by a factor of 20,  the new $\beta_{\rm data} = -2.98$  would imply that  $w$CDM is more likely to be the true model of the Universe than $\Lambda$CDM (see Fig. \ref{fig:betaSN_future}). Finally, Fig. \ref{fig:pridep1} highlights how the value of $\beta$ depends on the prior width. Relying on the Jeffreys scale in such cases might result in inaccurate predictions for the correct cosmological model.

\section{Conclusions}

The decisive role of statistics in testing physical models cannot be overestimated, especially in view of the very abundant data that are expected to be generated by current and future experiments. Having a number of alternative and complementary tools at our disposal to compare theoretical models is therefore a pressing issue. In this paper we argued that some shortcomings of standard Bayesian analysis can be overcome in the framework of a mixed frequentist/Bayesian approach, denoted as FB method, in which Bayes' ratio is employed as a frequentist statistic (see also \cite{Keeley:2021dmx} for a similar point of view). 
In particular, we have emphasized the dependence of Bayes' ratio on the prior, even in the limit of weak priors, and have shown that this issue is solved in our mixed approach.

The frequentist distribution of Bayes' factor can be employed both as a quality-of-fit test and as model comparison, thereby producing an exhaustive overview of competing models.
We also found the  analytical expression of such distribution in the case of correlated Gaussian data with linear nested models and arbitrary Gaussian priors, and shown that it provides a good approximation also in a realistic setting of interest to cosmology. In such case, the FB method  can be carried out without the need of data simulations.

In the case of non-Gaussian posteriors, the analytical results presented in this work are generally not applicable. Recent advances in simulation-based inference and emulators may, however, warrant an extension of the present method beyond the Gaussian approximation while still benefiting from its insight and fast applicability. Such an extension could involve building an emulator for the non-Gaussian correction to the FB distribution that learns from a (potentially small) number of simulations for which the Bayes factor is known exactly (e.g., computed using nested samplers such as {\tt PolyChord} \citep{1506.00171} or {\tt Nautilus} \citep{2306.16923}, or simulation-based methods, such as the one proposed in \cite{2207.04037}.)

\section*{Acknowledgments}

LA thanks  Juan Garcia-Bellido, Valerio Marra, Savvas Nesseris, and Bj\"orn M. Sch\"afer for useful discussions on these topics. ZS and LA acknowledge support from DFG project  456622116.  LA acknowledges support by the Deutsche Forschungsgemeinschaft (DFG, German Research Foundation) under Germany's Excellence Strategy EXC 2181/1 - 390900948 (the Heidelberg STRUCTURES Excellence Cluster). KW acknowledges support from the Science and Technology Facilities Council (STFC) under grant ST/X006344/1. More detailed calculation of several expressions can be found in the earlier versions of the arXiv submission.

\appendix

\section{A simple example}
\label{sec:simple}

\begin{figure}
\begin{centering}
\includegraphics[width=12cm]{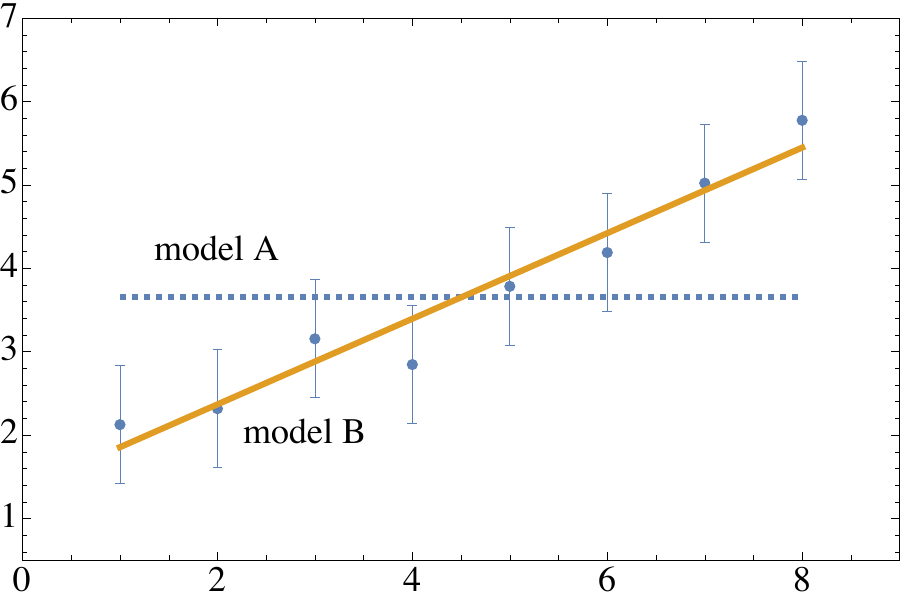}
\par\end{centering}
\caption{\label{fig:simple} This plot appear to show that model B is a better model than A in explaining the data. However, if the extra B parameter has an extremely weak prior, Bayes' factor $b$ would strongly favor model A.}
\end{figure}

Fig. \ref{fig:simple} shows an extremely simple example of how misleading weak priors can be. Here, model A is a horizontal straight line $y=a$ with just $a$ as free parameter, while model B is a general straight line $y=a+bx$ with two free parameters. The data points have been created by sampling a Gaussian distribution around the line $y=1+0.6x$, with uncertainty $\sqrt{1/2}$.  The blue dashed line is the best fit for model A, while the continuous orange line is the best fit for B.  The Gaussian prior standard deviation for the common parameter $a$ is 10, while for the extra parameter $b$ is huge, $10^6$. Clearly, model B appears strongly favoured by this data set. However, the very weak prior shifts $\beta$ to roughly 9.8, which corresponds to a Bayes' ratio $\approx 130$, well into the "decisive" range in favor of model A of the Jeffreys scale. So a pure Bayesian analysis of this simple case would produce the puzzling result that  model A is to be much preferred over B. Bayes' factor becomes smaller than unity (i.e., model B is favoured) only if the prior uncertainty on $b$ is smaller than 7000.

\bibliographystyle{unsrtnat}
\bibliography{syst-bias,references}

\end{document}